\documentclass[aps,preprint,groupedaddress,showpacs]{revtex4-1}

\usepackage[labelfont=bf,labelsep=period,justification=raggedright]{caption}
\usepackage{graphicx}
\usepackage{color}
\usepackage{comment}
\usepackage{subfig}
\usepackage{amsmath}
\usepackage{url}

\begin{document}


\title{Higher contagion and weaker ties mean \\anger spreads faster than joy in social media}


\author{Rui Fan\textsuperscript{1}, Ke Xu\textsuperscript{1} and Jichang Zhao\textsuperscript{2,*}}
\affiliation{$1$ State Key Lab of Software Development Environment, Beihang University\\
$^2$ School of Economics and Management, Beihang University \\
$^*$Corresponding author: jichang@buaa.edu.cn}


\date{\today}

\begin{abstract}
Increasing evidence suggests that, similar to face-to-face communications, human emotions also spread in online social media. However, the mechanisms underlying this emotional contagion, for example, whether different feelings spread in unlikely ways or how the spread of emotions relates to the social network, is rarely investigated. Indeed, because of high costs and spatio-temporal limitations, explorations of this topic are challenging using conventional questionnaires or controlled experiments. Because they are collection points for natural affective responses of massive individuals, online social media sites offer an ideal proxy for tackling this issue from the perspective of computational social science. In this paper, based on the analysis of millions of tweets in Weibo, surprisingly, we find that anger is more contagious than joy, indicating that it can spark more angry follow-up tweets. Moreover, regarding dissemination in social networks, anger travels easily along weaker ties than joy, meaning that it can infiltrate different communities and break free of local traps because strangers share such content more often. Through a simple diffusion model, we reveal that greater contagion and weaker ties function cooperatively to speed up anger's spread. The diffusion of real-world events with different dominant emotions provides further testimony to the findings. To the best of our knowledge, this is the first time that quantitative long-term evidence has been presented that reveals a difference in the mechanism by which joy and anger are disseminated. Our findings shed light on both personal anger management in human communications and on controlling collective outrage in cyberspace.
\end{abstract}


\maketitle


\section{Introduction}
\label{sec:intro}

Emotions have a substantial effect on human decision making and behavior~\cite{Gneezy28012014}. Emotions also transfer between different individuals through their communications and interactions, indicating that emotional contagion, which causes others to experience similar emotional states, could promote social interactions~\cite{Nummenmaa12062012,marsella2014computationally} and synchronize collective behavior, especially for individuals who are involved in social networks~\cite{Tadic2013} in the post-truth era~\cite{posttruth}. From this perspective, a better understanding of emotional contagion can disclose collective behavior patterns and help improve emotion management. However, the details of the mechanism by which emotional contagion spreads in a social network context remain unclear.

Conventional approaches such as laboratory experiments have been pervasively employed to attest to the existence of emotional contagion in real-world circumstances~\cite{hatfield1993emotional,fowler2008dynamic,rosenquist2011social}. However, unravelling the details of the mechanism behind emotional contagion is considerably more challenging because it is difficult for controlled experiments to establish large social networks, stimulate different emotions among the members of the network simultaneously, and then, track the propagation of emotions in real-time for long-term experiments. Meanwhile, to study the relationship dynamics between social structure and emotional contagion, properties such as the strength of relationships should also be considered; however, such considerations might introduce uncontrolled contextual factors that fundamentally undermine the reliability of the experiment. We argue that it is extremely difficult for conventional approaches to investigate the mechanism by which emotion spreads in the context of large social networks and long-term observations and that the traces of natural affective responses from the massive numbers of individuals in online social media offer a new and computationally suitable perspective~\cite{Lazer721,Golder1878,marsella2014computationally}.

It is not easy to differentiate between online interactions and face-to-face communications in terms of emotional contagion~\cite{cmc_emotion}. However, increasing evidence from both Facebook and Twitter in recent years has consistently demonstrated the existence of emotional contagion in online social media~\cite{guillory2011upset,coviello2014detecting,gruzd2011happiness,chmiel2011collective,kramer2012spread,dang2012impact,kramer2014experimental}. Kramer et al. provided the first experimental evidence of emotional contagion in Facebook by manipulating the amount of emotional content in people's News Feeds~\cite{kramer2014experimental}. Later Ferrara and Yang revealed evidence of emotional contagion in Twitter~\cite{ferrara2015measuring}; however, instead of controlling the content, they measured the emotional valence of user-generated content and showed that posting positive and negative tweets both follow paths that garner significant overexposure, indicating the spread of different feelings. Evidence from both studies suggests that---even in the absence of the non-verbal cues inherent to face-to-face interactions~\cite{ferrara2015measuring}, emotions that involve both positive and negative feelings can still be transferred from one user to others in online social media. In fact, through posting, reposting and other virtual interactions, users in online social media express and expose their natural emotional states into the social networks in real-time as the context evolves. Hence, serving as a ubiquitous sensing platform, online social media collects emotional expression and captures its dissemination in the most realistic and comprehensive circumstances, thus providing us with an unprecedented proxy for obtaining universal insights into the underlying mechanisms of emotional contagion in social networks.

However, in existing studies, emotions in online social media are usually simplified into either positive or negative~\cite{bollen2011happiness,bliss2012twitter,kramer2014experimental,ferrara2015measuring}, while many fine-grained emotional states---particularly negative ones such as anger and disgust---are neglected. In fact, on the Internet, negative feelings such as anger might dominate online bursts of communications about societal issues or terrorist attacks and play essential roles in driving collective behavior during the event propagation~\cite{chmiel2011negative,Zhao2012}. Aiming to fill this critical gap, in this paper, we categorize human emotions
 into four categories~\cite{Zhao2012,fan2014anger}, joy, anger, disgust and sadness, and then attempt to disclose the mechanisms underlying their spread. Investigating fine-grained emotional states enriches the landscape of emotional contagion and makes it possible to systematically understand the mechanisms that underlie the relationships between social structures and propagation dynamics.

In this paper, we employ over 11 million tweets posted by approximately 100,000 users over half a year on Weibo (a variant of Twitter in China) to perform a computational and quantitative investigation of emotional contagion. Surprisingly, we found that among the four emotions we study, joy and anger demonstrate significant evidence of contagion, and anger is the most contagious emotion, indicating that, over the short term, given the same exposure levels, angry messages spark more follow-up tweets or retweets than do joyous messages. Besides, from the view of coupled dynamics with the social structure, different measures of relationship strengths consistently show that anger is disseminated through weaker ties than joy, indicating that angry tweets can break out of local traps and has a higher probability of achieving global coverage. We conjecture that this greater contagion and weaker ties can result in anger spreading faster than joy in online social media. Simulations using a diffusion model and over 40 million tweets from 616 online communication bursts in China further solidify this argument. During these sudden events, we find that negative messages in which anger is the dominating emotion achieve diffusion peaks at shorter intervals and at higher velocities than their joyful counterparts. Moreover, the model also discloses that greater contagion and weaker ties function collaboratively in boosting anger's spread: being more contagious increases the infection rate, and anger's preference for weak ties attracts a greater number of susceptible subjects by breaking free of local traps. Our findings about emotional contagion will be insightful for personal anger management and collective behavior understanding in realistic scenarios either online or offline.

\section{Results}
\label{sec:res}

For this study, we collected 11,753,609 tweets posted by 92,176 users from September 2014 to March 2015, including the following networks of these users. Through a Bayesian classifier trained in~\cite{Zhao2012}, each emotional tweet in this data set was automatically classified as expressing joy, anger, disgust or sadness. Then, the contagion tendency and structure preferences are accordingly investigated in the social network, which comprises approximately 100,000 subjects. Finally, a toy model and the realistic bursts in Weibo were employed to investigate the conjecture that anger spreads faster than joy. Using this model, 40,005,242 tweets concerning 616 different events were employed to perform the computational analysis~\footnote{All the data sets are publicly available at \protect \url{https://dx.doi.org/10.6084/m9.figshare.4311920}.}.

\subsection*{Anger is more contagious than joy}
\label{ssec:contagion}

Rather than manipulating the content users received~\cite{kramer2012spread}, we measured emotional contagion by analyzing the extent to which users are influenced emotionally by the tweets they receive in Weibo over a certain period (i.e., an observation window). Similar to~\cite{ferrara2015measuring}, we assumed that the emotions to which users have been exposed in the recent past will elicit similar emotional states and will be reflected by users' subsequent tweets. Each individual in Weibo is embedded into an online social network where the nodes represent users and the directed links represent following relationships. For example, when user $u$ (the follower) follows user $w$ (the followee), a tie is established from $u$ to $w$, indicating that tweets posted by $w$ will be pushed to $u$ in real-time. Given an observation window of $\Delta$ hours, a vector $v(t_u)=(p_{anger}, p_{disgust}, p_{joy}, p_{sadness})$ represents the emotion distribution of the tweets to which $u$ was exposed within $\Delta$ before posting the tweet $t_u$. Note that we omit the tweets whose total exposure was less than 20 tweets to ensure the reliability of the measurement. By averaging over each dimension of all tweets' exposure vectors, a vector $v_{all}$ can be obtained to reflect the baseline of emotion distribution before any tweet is posted in Weibo. While for a certain emotion $i$ ($i=1,2,3$ or $4$), by averaging over each dimension of all tweets with emotion $i$, a vector $v_i$ can be similarly obtained to represent the average emotion distribution before any tweet with emotion $i$ is posted in Weibo. Then, for emotion $i$, the difference (denoted as $d_i$) on dimension $i$ between $v_{all}$ and $v_i$ can intuitively reflect the contagion significance of $i$ in Weibo. A higher $d_i$ value implies that emotion $i$ will stimulate tweets with the same emotion more significantly in the near future at above-average levels, suggesting that its contagion is more evident from the perspective of emotion correlation.

\begin{figure}
\centering
\includegraphics[width=\linewidth]{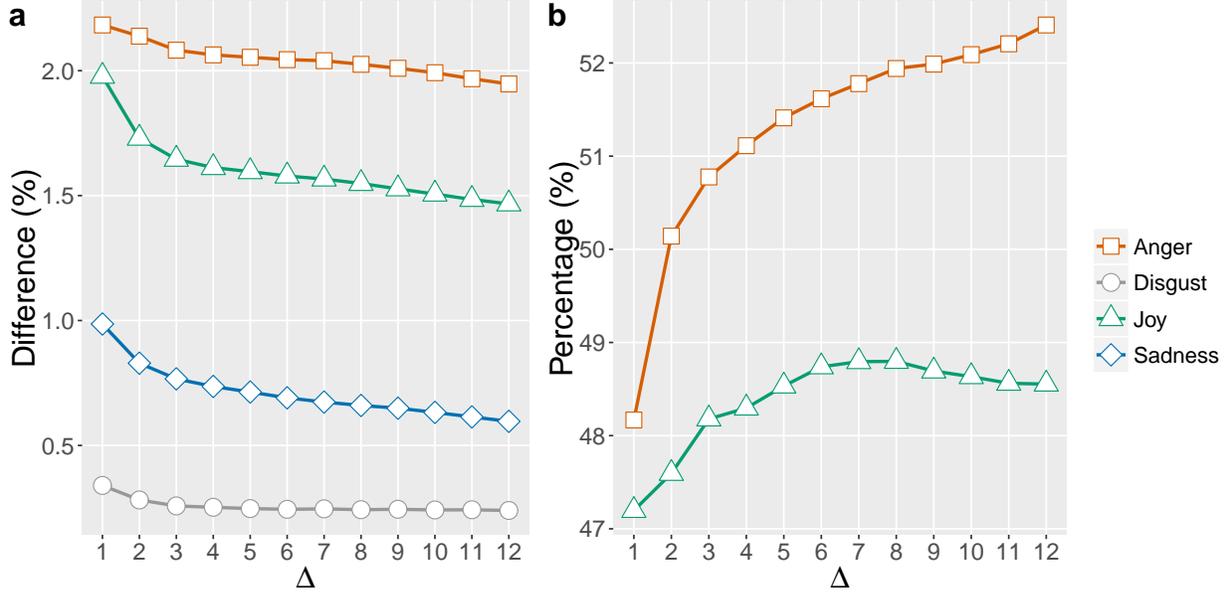}
\caption{Contagion significance and contagion tendency of different emotions. (a) The contagion significance of the various emotions declines with the length of observation window (in hours). (b) Contagion tendency, i.e., the percentage of influenced tweets for both anger and joy rises with the growth of the observation window and the fraction of anger-influenced tweets exceeds 50\% when $\Delta>4$ hours.}
\label{fig:contagion}
\end{figure}

As shown in Fig.~\ref{fig:contagion}(a), $d_{anger}$ and $d_{joy}$ are significantly higher than those of other emotions, which indicates that both emotions will spread evidently in social networks. In particular, anger possesses the highest significance compared to the baseline. For example, $d_{anger}$ is about 2.1\% when $\Delta=1$, which means that users received 2.1\% more angry tweets than usual in the time before they post an angry message. However, the probability of contagion is relatively low for disgust and sadness, especially for disgust, suggesting that the possibility of contagion for these emotions is trivial. Therefore, in the remainder of this paper, we focus only on anger and joy. Note that $d_i$ also decreases as $\Delta$ grows, because the influence of emotion decays with time; thus, users are more easily influenced by more recent messages, which is consistent with our intuition. As $\Delta$ grows, anger persistently shows the most significant difference, indicating that, as compared to other emotions, users' angry feelings are more likely to be simulated by their exposure to angry tweets in social media.

Emotion influence is closely entangled with emotional contagion. When emotions spread in a social network, connected individuals can influence each other emotionally, which might eventually lead to clusters of emotion (i.e., networked individuals demonstrate homophily in their sentimental states). For example, Bollen et al. found that users' happiness is assortative across Twitter~\cite{bollen2011twitter} and Bliss et al. revealed that average happiness scores are positively correlated between friends within three hops~\cite{bliss2012twitter}. In particular, Fan et al. showed that, in Weibo, anger is more strongly correlated than joy among friends within three hops, suggesting that anger is more influential than joy~\cite{fan2014anger}. Inspired by this, and similar to the metric presented in~\cite{ferrara2015measuring}, here, we further define the contagion tendency of anger and joy as the ratio of influenced tweets from the perspective of emotional influence to reflect the probability of contagion. For each tweet $t_u$ with emotion $i$ posted by $u$, we calculate the Euclidean distance $e_j$ ($j=1, 2, 3 \text{ or }4$) between $v(t_u)$ and $v_j$ ($j=1, 2, 3 \text{ or }4$). We define $t_u$ as an emotionally influenced tweet, i.e., $t_u$ is stimulated by the emotion $i$ carried by tweets that $u$ received within $\Delta$ when the emotion with the smallest distance is $i$. Then for anger and joy, the emotion with the highest percentage of influenced tweets is more influential and more contagious. As shown in Fig.~\ref{fig:contagion}(b), anger accrues a higher percentage of influenced tweets than joy, indicating that---compared to joy---users' anger is more easily stimulated by exposure to angry tweets. Meanwhile, as $\Delta$ grows, the influenced percentages of both emotions increase because users are exposed to more tweets when $\Delta$ is large. However, anger persistently achieves a higher percentage than joy and the gap increases as $\Delta$ grows longer. 

\begin{figure}
\centering
\includegraphics[width=\linewidth]{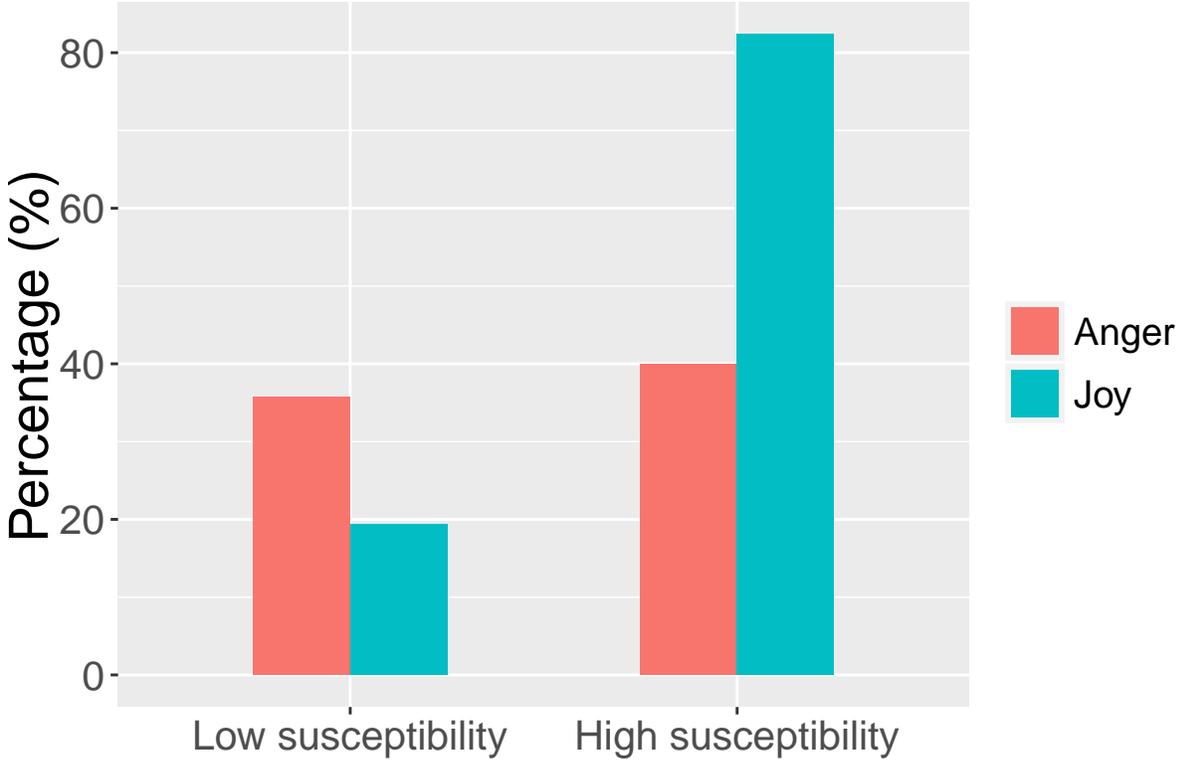}
\caption{The influenced percentage of anger and joy in users with high and low susceptibility.}
\label{fig:contagion_users}
\end{figure}

We further investigate emotional contagion from the viewpoint of sentimentally influenced users. Specifically, all users are sorted by the percentage of their influenced tweets in descending order (i.e., the fraction of influenced tweets over their tweeting timelines). The top and bottom 15\% of users are then defined as having high or low susceptibility, respectively. As can be seen in Fig.~\ref{fig:contagion_users}, users with low susceptibility are more easily affected by anger while users with high susceptibility are more inclined to be affected by joy. In contrast to ~\cite{ferrara2015measuring}, which claimed that both users with high and low susceptibility are more susceptible to positive feelings than to negative feelings, our results demonstrate that users with low susceptibility are more susceptible to anger---one of the negative emotions---than to joy. This implies that a fine-grained emotion category can lead to new insights when discussing emotional contagion and that anger, unexpectedly, is more contagious than joy. Meanwhile, the relatively smaller numbers of followers of users with low susceptibility 
indicates that anger has a greater influence on people with low susceptibility than on those with high susceptibility in social networks and that, therefore, given the power-law like distribution of follower numbers, anger is more contagious for the majority of social network users than is joy. 

Being more contagious and the fact that the majority are more susceptible to it indicate that anger can be transferred faster than joy, because it will spark more follow-up tweets under the same stimulus intensity. However, the mechanism underlying emotional contagion also relies heavily on social network structure and, thus, how structure functions in the contagion of joy and anger deserves further exploration.

\subsection*{Anger prefers weaker ties than joy}
\label{subsec:ties}

The dynamics of emotional contagion are essentially coupled with the underlying social network, which provides channels for disseminating a sentiment from one individual to others. Other than the willingness to post an angry or joyful tweet after having been exposed to anger or joy from followees, another key factor that determines users' actions is the relationships through which contagion is more likely to occur. Specifically, being able to predict which friend would be most likely to retweet an emotional post in the future would be a key insight for contagion modeling and control. 

Social relationships, or ties in social network, can be measured by their strength. Such measurements are instrumental in spreading both online and real-world human behaviors~\cite{Bond2012}. The strength of a tie in online social media can be intuitively measured using online interactions (i.e., retweets) between its two ends. Here, we present three metrics to depict tie strengths quantitatively. The first metric is the proportion of common friends~\cite{onnela2007structure,Zhao2010}, which is defined as $c_{ij}/(k_i-1+k_j-1-c_{ij})$ for the tie between users $i$ and $j$, where $c_{ij}$ denotes the number of friends that $i$ and $j$ have in common, and $k_i$ and $k_j$ represent the degrees of $i$ and $j$, respectively. Note that for the metric of common friends, the social network of Weibo is converted to an undirected network where each link represents possible interactions between both ends. The second metric is inspired by reciprocity in Twitter-like services: a higher ratio of reciprocity indicates more trust and more significant homophily~\cite{zhu2014influence}. Hence, for a given pair of users, the proportion of reciprocal retweets in their total flux is defined as the tie strength. The third metric is the number of retweets between two ends of a tie in Weibo, called retweet strength: larger values represent more frequent interactions. Note that, different from the previous two metrics, retweet strength evolves over time; therefore, here, we count only the retweets that occurred before the relevant emotional retweet. Moreover, to smooth the comparison between anger and joy, the retweet strength (denoted as $S$) is normalized by $(S-S_{min})/(S_{max}-S_{min})$, in which $S_{min}$ and $S_{max}$ separately represent the minimum and maximum values of all observations.

\begin{figure}
\centering
\includegraphics[width=\linewidth]{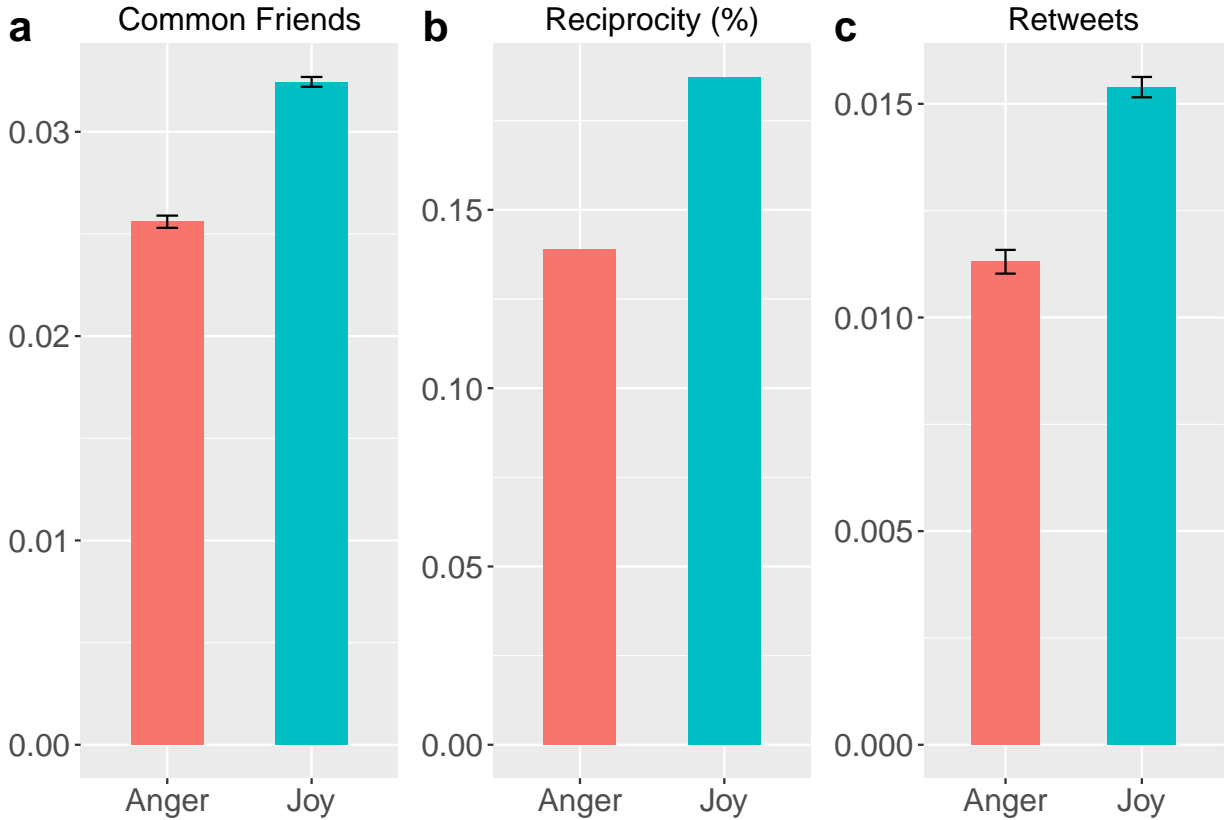}
\caption{Anger prefers weaker ties than joy. Three different metrics are averaged over all the emotional retweets in the dataset. The lower metrics for anger suggest that, in contagion, anger disseminates through weaker ties than joy. The error bars represent standard errors in (a) and (c), while in (b) there is no standard error because the reciprocity is simply a ratio obtained from all emotional retweets.}
\label{fig:weak_tie}
\end{figure}

\begin{figure*}
\centering
\includegraphics[width=0.8\linewidth]{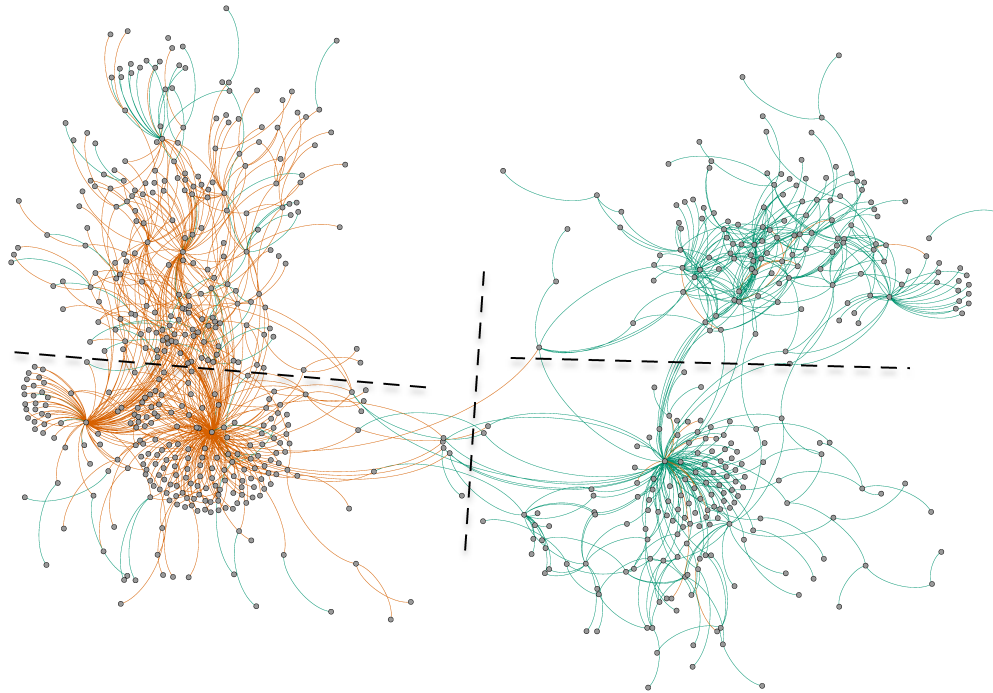}
\caption{Emotional contagion in a sampled snapshot of the Weibo network with four communities. Links with more angry retweets are orange; otherwise, they are green. It can be seen that anger prefers more weak ties that bridge different communities than does joy. The two communities to the left are dominated by anger; thus, a large percentage of their messages disseminate between communities. In contrast, the emotion of the two joy-dominated communities on the right tends to diffuse within the communities.}
\label{fig:graph}
\end{figure*}

By investigating each emotional retweet (i.e., a repost of an angry or joyful tweet posted by a followee), we can correlate emotional contagion with tie strengths. As shown in Fig.~\ref{fig:weak_tie}, all the metrics consistently demonstrate that anger prefers weaker ties in contagion than does joy, suggesting that angry tweets spread through weak ties with greater odds than do joyful tweets. It is well known that weak ties play essential roles in diffusion in social networks~\cite{granovetter1973strength}, particularly in breaking out of local traps caused by insular communities by bridging different clusters~\cite{onnela2007structure,Zhao2012KAIS,Meo2014}. For example, a typical snapshot of emotional contagion with four communities is illustrated in Fig.~\ref{fig:graph}, in which anger disseminates through weak ties of inter-communities more often than does joy. Because of this, compared to joy, anger has more chances to infiltrate different communities during emotional contagion because of its preference for weak ties. The increased number of infected communities leads to more global coverage, indicating that anger can achieve broader dissemination than joy over a given time period. 

The evidence from both contagion tendency and relationship strength suggests that anger can spread faster than joy because it will stimulate more follow-up tweets and infiltrate more communities in the Weibo social network. Next, a simple dissemination model and direct evidence from online communication bursts in Weibo will be presented to provide further systematic support for this conjecture.

\subsection*{Anger spreads faster than joy}
\label{subsec:fast}

No moods are created equally online~\cite{Choudhury2012}; the differences in contagion of positive and negative feelings has been a trending---but controversial---topic for years. Berger and Milkman revealed that positive content is more likely to become viral than negative content in a study of the NY Times~\cite{Berger2010}. Tadić and \v{S}uvakov found that, for human-like bots in online social networks, positive emotion bots are more effective than negative ones~\cite{Tadic2013}. Wu et al. even claimed that bad news containing more negative words fades more rapidly in Twitter~\cite{Wu2011}. In contrast, Chmiel et al. pointed out that negative sentiments boost user activity in BBC forums~\cite{chmiel2011negative} and Pfitzner et al. stated that users tend to retweet tweets with high emotional diversity~\cite{pfitzner2012emotional}. Ferrara and Yang demonstrated that although people are more likely to share positive content in Twitter, negative messages spread faster than positive ones at the content level~\cite{Ferra2015PJ}. Thus, Hansen et al. concluded that the relationship between emotion and virality is quite complicated~\cite{Hansen2011}. Here, we argue that a finer-grained division of human emotion, especially among negative emotions, will enrich the background for investigating contagion differences. In the meantime, creating explicit definitions of what fast spread entails will further facilitate the elimination of debate on this issue. 

To comprehensively understand how contagion tendency and tie strength function in the dynamics of emotion spread, we first establish a toy model to simulate such diffusion in Weibo's undirected following graph. In this model, which is inspired by the classic Susceptible-Infected (SI) model, first, a random seed with a certain emotion is selected to ignite the spread, and then, other susceptible nodes with no emotion will become infected and acquire the same sentimental state of the seed. Any susceptible node $s$ with an infected neighbor $i$ will become infected with the probability $p=\gamma~w_{is}^{\alpha}/{\sum_n{w_{in}^{\alpha}}}$, where $\gamma \ge 0$ reflects the contagious tendency, $\alpha$ controls the relationship strength preference and $n$ is one of $i$'s neighbors. Specifically, a greater $\gamma$ suggests that the emotion is more contagious, an $\alpha <0$ will preferentially select weak ties to spread the emotion, while strong ties are more likely to be selected when $\alpha > 0$, and $\alpha=0$ will cause a random selection of the diffusion path.

\begin{figure*}
\centering
\includegraphics[width=0.8\linewidth]{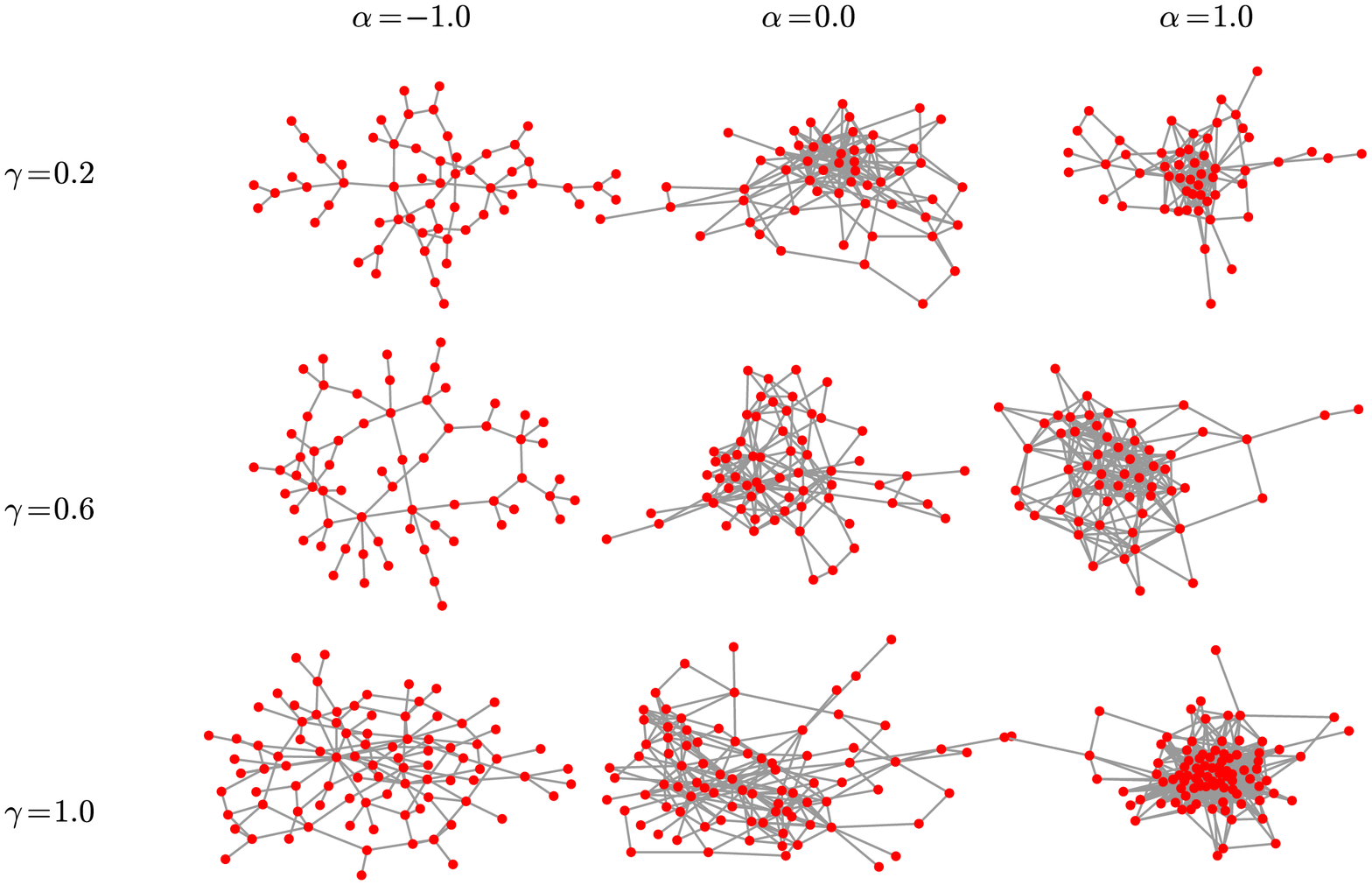}
\caption{Snapshots of the first infected 50 nodes in Weibo. Starting from the same seed, diverse values of $\gamma$ and $\alpha$ are used to simulate the spread. Note that in these simulations, the Weibo following graph was converted to an undirected graph.}
\label{fig:toy_model_snapshots}
\end{figure*}

As can be seen in Fig.~\ref{fig:toy_model_snapshots}, for a fixed $\gamma$, the infected networks produced by $\alpha=-1$ possess larger diameters than is the case for $\alpha=1$, and for a fixed $\alpha$, a greater $\gamma$ always leads to a more dense set of connections locally. Consistent with our previous conjecture, the preliminary observations above promisingly imply that, compared to strong ties, adopting weak ties as the diffusion paths improve the chances that an emotion will penetrate other network components, which then results in a large diameter and spare structures. In the meantime, a greater $\gamma$ means that more neighbors will be infected, which results in a dense neighborhood. By explicitly defining the speed of emotion spread, we next try to disclose which configurations of $\alpha$ and $\gamma$ will result in the fastest propagation. 

\begin{figure}
\centering
\includegraphics[width=\linewidth]{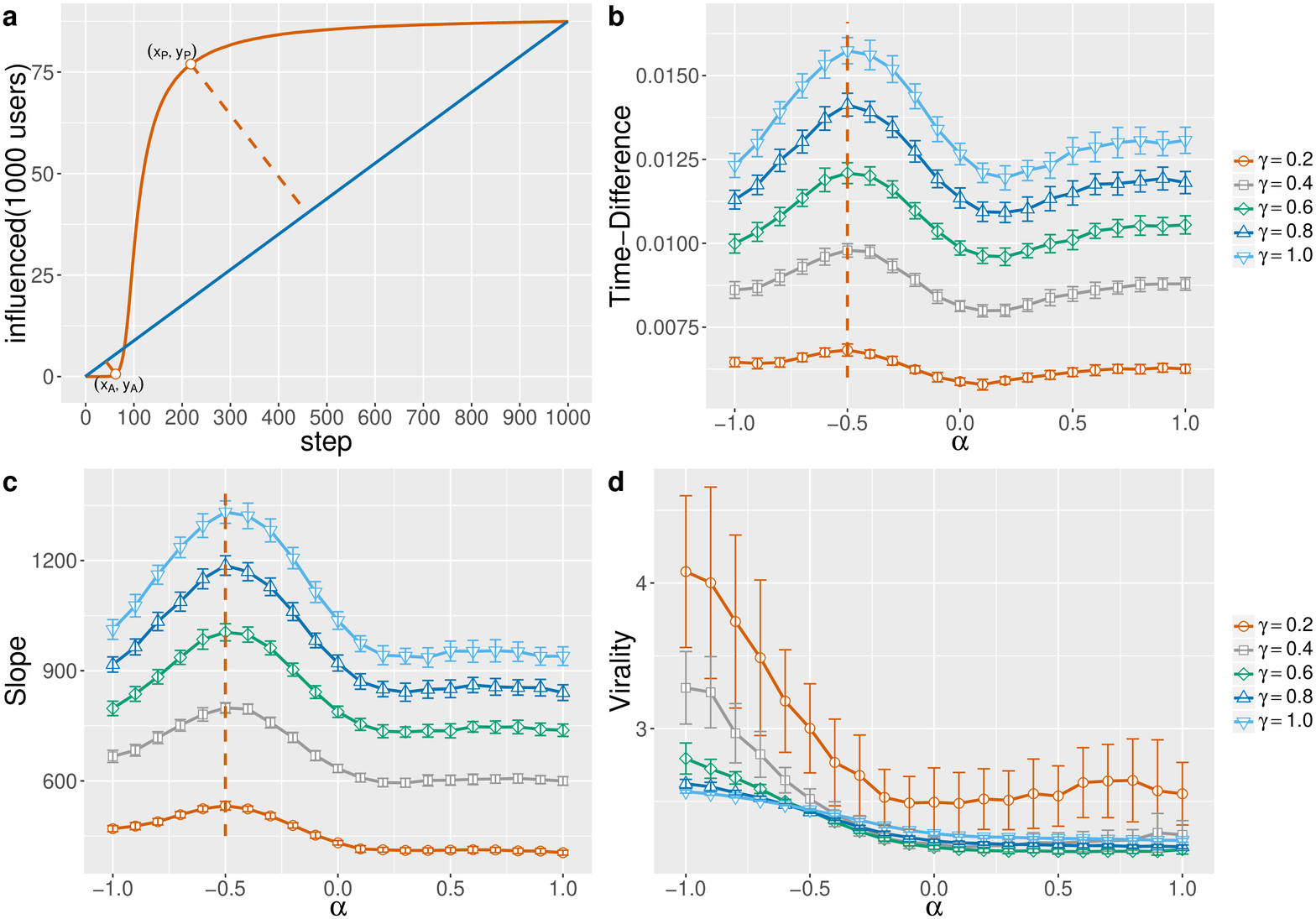}
\caption{Spread velocities and structures of the model. (a) The diagrammatic volume of infected nodes grows with time, where $x_A$ denotes the instant of awakening, $x_P$ denotes the peak time and $y_A$ and $y_P$ stand for the volume of infected nodes at the instants of awakening and peak, respectively. The first and last points of the accumulating curve are connected to obtain the reference line $L$. Then, the peak instant can be identified as the point the furthest distance from $L$ and located above $L$. The awakening instant can be found using the same method, except that the point will be located under $L$. (b) The time-difference varies with $\alpha$ for different $\gamma$ values. (c) The slope varies with $\alpha$ for different $\gamma$ values. (d) The virality of the diffusion structure varies with $\alpha$ for different $\gamma$ values. All the simulations were performed on the undirected following graph of Weibo. For each pair of $\alpha$ and $\gamma$ values, 50 simulations with random seeds were conducted to guarantee stable statistics. The standard deviations are also presented.}
\label{fig:toy_model}
\end{figure}

For online emotion spread, ``fast spread'' means that the volume of emotional tweets grows quickly during the period from the awakening instant to the diffusion peak. Through a parameterless approach presented in~\cite{ke2015defining}, we can precisely locate the instants of both awakening and peak. Then, the spread speed is reflected by the interval between the awakening and the peak (shorter intervals denote faster spread) or by the average velocity of the growth in diffusion from awakening to peak (higher velocities indicate faster spread). Specifically, as demonstrated in Fig.~\ref{fig:toy_model}(a), the interval $1/(x_P-x_A)$, denoted as the time-difference, reflects the speed of the spread, while the average velocity can be defined as the slope (i.e., $(y_P-y_A)/(x_P-x_A)$). For both measures, higher values indicate faster spread. 

As shown in Fig.~\ref{fig:toy_model}(b) and (c), the speed of emotion spread from the perspective of both time-interval and velocity reaches its maximum as $\alpha \approx -0.5$, and larger $\gamma$ values generate higher maximum speed values. It can be concluded that when $\alpha<0$, preferentially selecting weak ties as the diffusion path can greatly boost the velocity of the spread. Being more contagious will also result in a faster spread velocity. Meanwhile, the virality, defined as the average length of the shortest path among infected nodes, is an indicator that reflects how viral a message is in online social media~\cite{structural_virality} and can be employed to study the structural preferences of the different model settings. As shown in Fig.~\ref{fig:toy_model}(d), in our toy model, the emotion's virality declines rapidly as $\alpha$ grows from -1 to 1 and, surprisingly, higher $\gamma$ values lead to less virality. Recall the spread snapshots from Fig.~\ref{fig:toy_model_snapshots}. Here, an $\alpha<0$ is inclined to select weak ties that help emotion penetrate the more distant parts of the network, leading to a larger diameter and thus higher virality. However, a large $\gamma$ makes the neighbors more likely to become infected, creating dense local structures during the spread, which tend to trap the emotion locally, decrease the length of the shortest path and, accordingly, reduce virality. Hence greater contagion and weaker ties function collaboratively to boost emotion spread. In other words, greater contagion infects more neighbors, but weaker ties infiltrate more communities, breaking out of local traps. Jointly, greater contagion and weaker ties act to boost the velocity of the emotion spread. 

To summarize, the presented toy model clearly discloses the roles of greater contagion and weaker ties in emotion spread, revealing that they function in a cooperative manner in which weak ties help emotion reach more susceptible individuals and greater contagion increases the infection rate. The simulations thoroughly support our conjecture that greater contagion and weaker ties will lead to the faster spread of anger compared to joy. This is particularly true of the weak ties, which were conventionally thought to act as bridges only for novel information~\cite{granovetter1973strength} instead of for negative feelings such as anger. Additional evidence from real-world emotion spread is presented in the next section to support our explanations. 

\begin{figure}
\centering
\includegraphics[width=\linewidth]{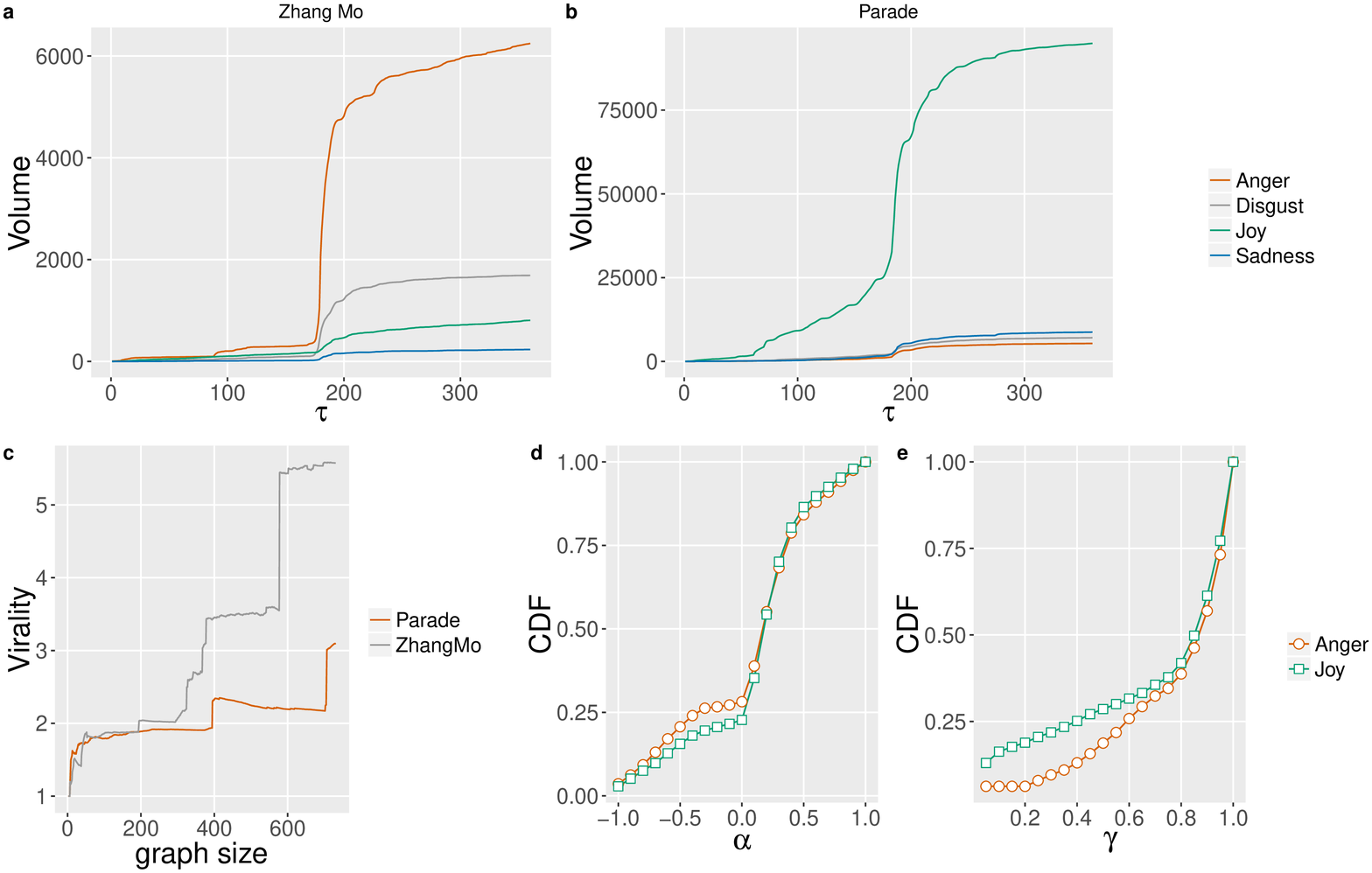}
\caption{Evidence from realistic online communication bursts. (a) An example of an anger-dominated event (celebrity scandal) in which cumulatively posted emotional tweets vary over time $\tau$. (b) An example of a joy-dominated event (a National Holiday parade). (c) Virality comparison between an anger-dominated event (Zhang Mo) and a joy-dominated event (Parade). Here the network of infected nodes was extracted from the retweeting graph instead of the following graph, which was missing from our data sets. Note that the virality was averaged over the isolated components that constitute the retweeting graph. (d) The cumulative distribution function (CDF) of $\alpha$ for online communication bursts. (e) The CDF of $\gamma$ for online communication bursts. Note that for (d) and (e), the cumulative volume of each online burst was normalized by dividing its peak ($y_P$) and the 20 most fitting pairs of $\alpha$ and $\gamma$ for each event were selected to produce the statistics.}
\label{fig:event}
\end{figure}

Over 600 communication burst events were extracted from Weibo. For each event, the emotion that occupied more than 60\% emotional tweets was defined as the dominant emotion of the event. In total we obtained 37 anger-dominated events and 200 joy-dominated events, as shown in Fig.~\ref{fig:event}(a), in which the cumulative volume of emotional tweets grows and then becomes saturated with time, at which point anger takes over the majority of the tweets. Here, the time granularity of all burst events is one hour. Several events with no obvious awakening or peak instants were omitted. 
It is worth noting that the small number of anger-dominated events explains its low prevalence in Weibo~\cite{Zhao2012}. It is also interesting that with respect to the virality of the events shown in Fig.~\ref{fig:event}(c), the anger-dominated event demonstrates the highest virality, especially as the retweeting graph grows, suggesting that weak ties are preferred in real-world anger dissemination and that the $\alpha$ parameter of the toy model should be set to a negative value to model angry bursts. Considering the definition of the spread velocity, we fit the toy model to the realistic events only in the range from the awakening instant to the peak timing. By employing the Dynamic Time Warping (DTW)~\cite{Berndt1994dtw} approach, the combinations of $\alpha$ and $\beta$ that resulted in the smallest distances between the simulation and the real-world results were selected as the optimal parameters to fit the online bursts. As shown in Figs.~\ref{fig:event} (d) and (e), compared to the joy-dominated events, events where $\alpha<0$ and large $\gamma$ ($\ge 0.8$) occupy a larger portion of the anger-dominated events, implying that in real-world online bursts, higher contagion and weaker ties indeed contribute to the spread of anger, and---as we revealed in the model---they may also increase the velocity of the angry bursts. The average results of spread velocity, i.e., the time-difference and slope, for all anger- and joy-dominated events are listed in Table~\ref{tab:speed}, where it can be seen that both metrics significantly support the simulation results that anger-dominated events tend to arrive at the peak from the awakening more quickly than do joy-dominated events. We can conclude, based on evidence from both model simulations and real-world events, that anger indeed spreads faster than joy in social media such as Weibo. 

\begin{table}[htbp]
\centering
\begin{tabular}{lll}
\hline
\textbf{Measures} & \textbf{Anger} & \textbf{Joy} \\
\hline
Time-difference &  0.018    &   0.014\\
Slope   & 0.027     &   0.020\\
\hline
\end{tabular}
\caption{Averaged values of all anger- and joy-dominated events. Note that here the slope is normalized by dividing the peak volume of events, i.e., $y_P.$}
\label{tab:speed}
\end{table}

\section{Discussion}
\label{sec:dis}

It has been said that bad is always stronger than good~\cite{bad_is_stronger}. Previous studies clearly show that, in Weibo, anger is more influential than joy~\cite{fan2014anger} and that, in Twitter, negative content spreads faster than positive content~\cite{Ferra2015PJ}. However, in this study, by finely splitting negative emotions into anger, disgust and sadness, we are the first to offer empirical evidence that anger spreads faster than joy in social media such as Weibo. As written in an ancient Chinese book entitled ``Zhongyong (The Doctrine of the Mean)'' more than two thousand years ago, ``when the feelings (e.g. pleasure, anger, sadness, or joy) have been stirred, and they act in their due degree, there ensues what may be called the state of HARMONY.'' Our study implies that we should place a stronger emphasis on anger in both personal emotional management and in collective mood understanding to make social media reach this state.

The way anger is expressed and experienced on the Internet is gaining attention. Rant-site visitors have self-reported that posting angry messages produces immediate feelings of relaxation~\cite{martin2013anger}, which makes posting anger an effective method for self-regulating mood. However, considering the high contagion of anger, an increase in angry expressions on the Internet might arouse negative shifts in the mood of the crowds connected to the posters~\cite{martin2013anger,Park2012}. Moreover, anger's preference for weak ties make it more likely to spread to ``strangers'' on the
 Internet. Users who want to assuage their anger by posting on the Internet should be made to understand the possible impact that such posts can have on their online social networks. Even in offline scenarios such as the ``road rage''~\cite{roadrage} that occurs during rush hours in China, anger among strangers can spread quickly and cause aggressive driving or even accidents. We suggest that in personal anger management, the possibility of infecting strangers---however unexpected--should be seriously considered. 

Online social media has become the most ubiquitous platform for collective intelligence, in which various signals generated by the massive numbers of connected individuals provide a foundation for understanding collective behavior. It is even possible that in the post-truth era, public opinion might be driven or shaped by emotional appeals rather than objective facts~\cite{posttruth}. However, how emotional contagion affects individual behavior in the aggregate---particularly individual intelligence---is rarely considered. In our findings, the spread of emotion, especially the high contagion and high velocity of anger, might have large implications concerning collective behavior in cyberspace, particularly with regard to crowdsourcing. It has even been stated that emotion such as anger can be a threat to reasoned argument~\cite{marsella2014computationally}. For example, outrage provoked in massive numbers of emotional individuals can profoundly bias public opinion, might originate with the fast spread of anger but not be due the event itself. Meanwhile, the high contagion and fast spread of anger also offers a new perspective for picturing the emotional behavior of crowds on the Internet. We suggest that in scenarios such as crowdsourcing and in understanding collective behavior, angry users should be treated carefully to negate or reduce their possible impact on the fair judgment of the observations in the big data era. In addition, reducing weak ties can function effectively in controlling the diffusion of Internet-fueled outrage and help make crowds more rational and smarter.

In contrast, happiness is believed to unify and cluster within communities~\cite{connected}. Our findings about joy's preferences for stronger ties supports this concept (see~Figs. \ref{fig:weak_tie} and \ref{fig:graph}), implying that the stronger ties inside communities are more suitable for disseminating joyful content in online social media. Meanwhile, self-reports from Facebook users also testify that communication with strong ties is associated with improvements in well-being~\cite{Burke2016}, which further indicates that our findings from the computational viewpoint are solid. 

\section{Conclusion}
\label{sec:con}

Rather than using self-reports in controlled experiments, this study collected natural and emotional postings from the Weibo social network to investigate the mechanism of emotional contagion in detail from the new viewpoint of computational social science. For the first time, we offer solid evidence that anger spreads faster than joy in social media because it is more contagious and disseminates preferentially through weak ties. Our findings shed light on both personal anger management and in understanding collective behavior. 

This study inevitably has limitations. It is generally accepted that emotion expression is culture dependent and that demographics such as gender also matter~\cite{association_theory,hu2016ambivalence}, suggesting that exploring how anger spreads in Twitter and how male and female responses differ regarding emotional contagion may be of great significance in future work.

\section*{Acknowledgment}
\label{sec:ack}

This work was supported by the National Natural Science Foundation of China (Grant Nos. 71501005, 71531001 and 61421003) and the fund of the State Key Lab of Software Development Environment (Grant Nos. SKLSDE-2015ZX-05 and SKLSDE-2015ZX-28). R. F. also thanks the Innovation Foundation of BUAA for PhD Graduates.



\begin{thebibliography}{10}

\bibitem{association_theory}
Richard~P. Bagozzi, Nancy Wong, and Youjae Yi.
\newblock The role of culture and gender in the relationship between positive
  and negative affect.
\newblock {\em Cogn Emot}, 13(6):641--672, 1983.

\bibitem{bad_is_stronger}
Roy~F. Baumeister, Ellen Bratslavsky, Catrin Finkenauer, and Kathleen~D. Vohs.
\newblock Bad is stronger than good.
\newblock {\em Rev Gen Psychol}, 5(4):323--370, 2001.

\bibitem{Berger2010}
Jonah Berger and Katherine~L. Milkman.
\newblock Social transmission, emotion, and the virality of online content.
\newblock Wharton Research Paper 106, 2010.

\bibitem{Berndt1994dtw}
Donald~J. Berndt and James Clifford.
\newblock Using dynamic time warping to find patterns in time series.
\newblock In {\em Proceedings of the 3rd International Conference on Knowledge
  Discovery and Data Mining}, AAAIWS'94, pages 359--370, 1994.

\bibitem{bliss2012twitter}
Catherine~A Bliss, Isabel~M Kloumann, Kameron~Decker Harris, Christopher~M
  Danforth, and Peter~Sheridan Dodds.
\newblock Twitter reciprocal reply networks exhibit assortativity with respect
  to happiness.
\newblock {\em J Comput Sci}, 3(5):388--397, 2012.

\bibitem{bollen2011happiness}
Johan Bollen, Bruno Gon{\c{c}}alves, Guangchen Ruan, and Huina Mao.
\newblock Happiness is assortative in online social networks.
\newblock {\em Artif Life}, 17(3):237--251, 2011.

\bibitem{bollen2011twitter}
Johan Bollen, Huina Mao, and Xiaojun Zeng.
\newblock Twitter mood predicts the stock market.
\newblock {\em J Comput Sci}, 2(1):1--8, 2011.

\bibitem{Bond2012}
Robert~M. Bond, Christopher~J. Fariss, Jason~J. Jones, Adam D.~I. Kramer,
  Cameron Marlow, Jaime~E. Settle, and James~H. Fowler.
\newblock A 61-million-person experiment in social influence and political
  mobilization.
\newblock {\em Nature}, 489(7415):295--298, 2012.

\bibitem{Burke2016}
Moira Burke and Robert~E. Kraut.
\newblock The relationship between {F}acebook use and well-being depends on
  communication type and tie strength.
\newblock {\em J Comput Mediat Commun}, 21:265–281, 2016.

\bibitem{chmiel2011collective}
Anna Chmiel, Julian Sienkiewicz, Mike Thelwall, Georgios Paltoglou, Kevan
  Buckley, Arvid Kappas, and Janusz~A Ho{\l}yst.
\newblock Collective emotions online and their influence on community life.
\newblock {\em PloS ONE}, 6(7):e22207, 2011.

\bibitem{chmiel2011negative}
Anna Chmiel, Pawel Sobkowicz, Julian Sienkiewicz, Georgios Paltoglou, Kevan
  Buckley, Mike Thelwall, and Janusz~A Ho{\l}yst.
\newblock Negative emotions boost user activity at {BBC} forum.
\newblock {\em Physica A Stat Mech Appl}, 390(16):2936--2944, 2011.

\bibitem{connected}
Nicholas~A. Christakis and James~H. Fowler.
\newblock {\em Connected: The Surprising Power of Our Social Networks and How
  They Shape Our Lives}.
\newblock Back Bay Books, New York, NY, 2011.

\bibitem{coviello2014detecting}
Lorenzo Coviello, Yunkyu Sohn, Adam~DI Kramer, Cameron Marlow, Massimo
  Franceschetti, Nicholas~A Christakis, and James~H Fowler.
\newblock Detecting emotional contagion in massive social networks.
\newblock {\em PloS ONE}, 9(3):e90315, 2014.

\bibitem{dang2012impact}
Linh Dang-Xuan and Stefan Stieglitz.
\newblock Impact and diffusion of sentiment in political communication-an
  empirical analysis of political weblogs.
\newblock In {\em Proceedings of the Sixth International AAAI Conference on
  Weblogs and Social Media}, pages 427--430, 2012.

\bibitem{Choudhury2012}
Munmun De~Choudhury, Scott Counts, and Michael Gamon.
\newblock Not all moods are created equal! {E}xploring human emotional states
  in social media.
\newblock In {\em Proceedings of the Sixth International AAAI Conference on
  Weblogs and Social Media}, ICWSM '12, pages 66--73, 2012.

\bibitem{Meo2014}
Pasquale De~Meo, Emilio Ferrara, Giacomo Fiumara, and Alessandro Provetti.
\newblock On {F}acebook, most ties are weak.
\newblock {\em Commun ACM}, 57:78--84, 2014.

\bibitem{posttruth}
Hossein Derakhshan.
\newblock Social media is killing discourse because it's too much like {TV}.
\newblock MIT Technology Review,
  \url{https://www.technologyreview.com/s/602981/social-media-is-killing-discourse-because-its-too-much-like-tv/},
  Nov 29 2016.

\bibitem{cmc_emotion}
Daantje Derks, Agneta~H. Fischer, and Arjan~E.R. Bos.
\newblock The role of emotion in computer-mediated communication: A review.
\newblock {\em Comput Hum Behav}, 24:766--785, 2008.

\bibitem{fan2014anger}
Rui Fan, Jichang Zhao, Yan Chen, and Ke~Xu.
\newblock Anger is more influential than joy: Sentiment correlation in {W}eibo.
\newblock {\em PLoS ONE}, 9:e110184, 2014.

\bibitem{ferrara2015measuring}
Emilio Ferrara and Zeyao Yang.
\newblock Measuring emotional contagion in social media.
\newblock {\em PloS ONE}, 10(11):e0142390, 2015.

\bibitem{Ferra2015PJ}
Emilio Ferrara and Zeyao Yang.
\newblock Quantifying the effect of sentiment on information diffusion in
  social media.
\newblock {\em PeerJ Comput Sci}, 1:e26, 2015.

\bibitem{fowler2008dynamic}
James~H Fowler and Nicholas~A Christakis.
\newblock Dynamic spread of happiness in a large social network: Longitudinal
  analysis over 20 years in the {Framingham Heart Study}.
\newblock {\em BMJ}, 337:a2338, 2008.

\bibitem{Gneezy28012014}
Uri Gneezy and Alex Imas.
\newblock Materazzi effect and the strategic use of anger in competitive
  interactions.
\newblock {\em Proc Natl Acad Sci USA}, 111(4):1334--1337, 2014.

\bibitem{structural_virality}
Sharad Goel, Ashton Anderson, Jake Hofman, and Duncan~J. Watts.
\newblock The structural virality of online diffusion.
\newblock {\em Manag Sci}, 62:180--196, 2016.

\bibitem{Golder1878}
Scott~A. Golder and Michael~W. Macy.
\newblock Diurnal and seasonal mood vary with work, sleep, and daylength across
  diverse cultures.
\newblock {\em Science}, 333(6051):1878--1881, 2011.

\bibitem{granovetter1973strength}
Mark~S Granovetter.
\newblock The strength of weak ties.
\newblock {\em Am J Sociol}, pages 1360--1380, 1973.

\bibitem{gruzd2011happiness}
Anatoliy Gruzd, Sophie Doiron, and Philip Mai.
\newblock Is happiness contagious online? {A} case of {T}witter and the 2010
  {Winter Olympics}.
\newblock In {\em 2011 44th Hawaii International Conference on System Sciences
  (HICSS)}, pages 1--9. IEEE, 2011.

\bibitem{guillory2011upset}
Jamie Guillory, Jason Spiegel, Molly Drislane, Benjamin Weiss, Walter Donner,
  and Jeffrey Hancock.
\newblock Upset now?: Emotion contagion in distributed groups.
\newblock In {\em Proceedings of the SIGCHI Conference on Human Factors in
  Computing Systems}, CHI '11, pages 745--748, New York, NY, USA, 2011. ACM.

\bibitem{Hansen2011}
Lars~Kai Hansen, Adam Arvidsson, Finn~Aarup Nielsen, Elanor Colleoni, and
  Michael Etter.
\newblock Good friends, bad news - affect and virality in {T}witter.
\newblock In James~J. Park, Laurence~T. Yang, and Changhoon Lee, editors, {\em
  Future Information Technology: 6th International Conference, FutureTech 2011,
  Loutraki, Greece, June 28-30, 2011, Proceedings, Part II}, pages 34--43.
  Springer Berlin Heidelberg, 2011.

\bibitem{hatfield1993emotional}
Elaine Hatfield, John~T Cacioppo, and Richard~L Rapson.
\newblock Emotional contagion.
\newblock {\em Curr Dir Psychol Sci}, 2(3):96--100, 1993.

\bibitem{hu2016ambivalence}
Yue Hu, Jichang Zhao, and Junjie Wu.
\newblock Emoticon-based ambivalent expression: A hidden indicator for unusual
  behaviors in {W}eibo.
\newblock {\em PLoS ONE}, 11:e0147079, 2016.

\bibitem{ke2015defining}
Qing Ke, Emilio Ferrara, Filippo Radicchi, and Alessandro Flammini.
\newblock Defining and identifying {Sleeping Beauties} in science.
\newblock {\em Proc Natl Acad Sci USA}, 112(24):7426--7431, 2015.

\bibitem{kramer2012spread}
Adam~D.I. Kramer.
\newblock The spread of emotion via {F}acebook.
\newblock In {\em Proceedings of the SIGCHI Conference on Human Factors in
  Computing Systems}, CHI '12, pages 767--770, New York, NY, USA, 2012. ACM.

\bibitem{kramer2014experimental}
Adam~DI Kramer, Jamie~E Guillory, and Jeffrey~T Hancock.
\newblock Experimental evidence of massive-scale emotional contagion through
  social networks.
\newblock {\em Proc Natl Acad Sci USA}, 111(24):8788--8790, 2014.

\bibitem{Lazer721}
David Lazer, Alex Pentland, Lada Adamic, Sinan Aral, Albert-L{\'a}szl{\'o}
  Barab{\'a}si, Devon Brewer, Nicholas Christakis, Noshir Contractor, James
  Fowler, Myron Gutmann, Tony Jebara, Gary King, Michael Macy, Deb Roy, and
  Marshall Van~Alstyne.
\newblock Computational social science.
\newblock {\em Science}, 323(5915):721--723, 2009.

\bibitem{marsella2014computationally}
Stacy Marsella and Jonathan Gratch.
\newblock Computationally modeling human emotion.
\newblock {\em Commun ACM}, 57(12):56--67, 2014.

\bibitem{martin2013anger}
Ryan~C Martin, Kelsey~Ryan Coyier, Leah~M Vansistine, and Kelly Schroeder.
\newblock Anger on the internet: The perceived value of rant-sites.
\newblock {\em Cyberpsychol Behav Soc Netw}, 16(2):119--122, 2013.

\bibitem{Note1}
All the data sets are publicly available at \protect \url
  {https://dx.doi.org/10.6084/m9.figshare.4311920}.

\bibitem{Nummenmaa12062012}
Lauri Nummenmaa, Enrico Glerean, Mikko Viinikainen, Iiro~P. Jääskeläinen,
  Riitta Hari, and Mikko Sams.
\newblock Emotions promote social interaction by synchronizing brain activity
  across individuals.
\newblock {\em Proc Natl Acad Sci USA}, 109(24):9599--9604, 2012.

\bibitem{onnela2007structure}
J-P Onnela, Jari Saram{\"a}ki, Jorkki Hyv{\"o}nen, Gy{\"o}rgy Szab{\'o}, David
  Lazer, Kimmo Kaski, J{\'a}nos Kert{\'e}sz, and A-L Barab{\'a}si.
\newblock Structure and tie strengths in mobile communication networks.
\newblock {\em Proc Natl Acad Sci USA}, 104(18):7332--7336, 2007.

\bibitem{Park2012}
Minsu Park, Chiyoung Cha, and Meeyoung Cha.
\newblock Depressive moods of users portrayed in {T}witter.
\newblock In {\em Proceedings of the ACM SIGKDD Workshop on Healthcare
  Informatics (HI-KDD)}, pages 1--8, 2012.

\bibitem{pfitzner2012emotional}
Ren{\'e} Pfitzner, Antonios Garas, and Frank Schweitzer.
\newblock Emotional divergence influences information spreading in {T}witter.
\newblock In {\em Proceedings of the Sixth International AAAI Conference on
  Weblogs and Social Media}, pages 2--5, 2012.

\bibitem{rosenquist2011social}
J~Niels Rosenquist, James~H Fowler, and Nicholas~A Christakis.
\newblock Social network determinants of depression.
\newblock {\em Mol Psychiatry}, 16(3):273--281, 2011.

\bibitem{Tadic2013}
Bosiljka Tadić and Milovan Šuvakov.
\newblock Can human-like bots control collective mood: Agent-based simulations
  of online chats.
\newblock {\em J Stat Mech Theory Exp}, 2013(10):P10014, 2013.

\bibitem{roadrage}
Wikipedia.
\newblock Road rage.
\newblock \url{https://en.wikipedia.org/wiki/Road_rage}, 2016.

\bibitem{Wu2011}
Shaowei Wu, Chenhao Tan, Jon Kleinberg, and Michael Macy.
\newblock Does bad news go away faster?
\newblock In {\em Proceedings of the Fifth International AAAI Conference on
  Weblogs and Social Media}, ICWSM '11, pages 646--649, 2011.

\bibitem{Zhao2012}
Jichang Zhao, Li~Dong, Junjie Wu, and Ke~Xu.
\newblock {MoodLens}: An emoticon-based sentiment analysis system for {Chinese}
  tweets.
\newblock In {\em Proceedings of the 18th ACM SIGKDD International Conference
  on Knowledge Discovery and Data Mining}, KDD '12, pages 1528--1531, New York,
  NY, USA, 2012. ACM.

\bibitem{Zhao2012KAIS}
Jichang Zhao, Junjie Wu, Xu~Feng, Hui Xiong, and Ke~Xu.
\newblock Information propagation in online social networks: A tie strength
  perspective.
\newblock {\em Knowl Inf Syst}, 32:589--608, 2012.

\bibitem{Zhao2010}
Jichang Zhao, Junjie Wu, and Ke~Xu.
\newblock Weak ties: Subtle role of information diffusion in online social
  networks.
\newblock {\em Phys Rev E}, 82:016105, Jul 2010.

\bibitem{zhu2014influence}
Yu-Xiao Zhu, Xiao-Guang Zhang, Gui-Quan Sun, Ming Tang, Tao Zhou, and Zi-Ke
  Zhang.
\newblock Influence of reciprocal links in social networks.
\newblock {\em PloS ONE}, 9(7):e103007, 2014.

\end{thebibliography}

\end{document}